\documentclass[aps,pra,twocolumn,showpacs,floatfix]{revtex4}

\usepackage{amsmath,graphicx}
\usepackage{dcolumn}
\begin{document}


\title{Dynamics of Photo-excited Spins in InSb Based Quantum Wells}

\author{ K. Nontapot, R. N. Kini, B. Spencer, G. A. Khodaparast}
\thanks{Author to whom correspondence should be addressed}
\email{khoda@vt.edu} \affiliation{Department of Physics, Virginia
Tech, Blacksburg, VA 24061}
\author{N. Goel, S. J. Chung, T. D. Mishima and, M. B. Santos}
\affiliation{Homer L. Dodge Department of Physics and Astronomy,
University of Oklahoma, Norman, OK, 73019}
\thanks{}

\date{\today}


\begin{abstract}
We report time resolved measurements of spin relaxation in doped and
undoped InSb quantum wells using degenerate and two-color
magneto-optical Kerr effect techniques. We observed that the
photo-excited spin dynamics are strongly influenced by laser
excitation fluence and the doping profile of the samples. In the low
fluence regime, an oscillatory pattern was observed at low
temperatures ($\leq$ 77 K) in the samples with an asymmetric doping
profile which might be attributed to the quasi-collision-free spin
relaxation regime. Our measurements also suggest the influence of
the barrier materials (Al$_{x}$In$_{1-x}$Sb) on the spin relaxation
in these material systems.
\end{abstract}


\maketitle
\section{Introduction}
In light of the growing interest in spin-related phenomena and
devices, there is now a renewed interest in the science and
engineering of narrow gap semiconductors (NGS) such as InSb. NGS
offer several scientifically unique electronic features such as a
small effective mass, a large g-factor, a high intrinsic mobility,
and large spin-orbit coupling effects. In semiconductors with large
spin-orbit interaction the coupling of electron spin polarization
with electric fields or currents can provide new opportunities for
spin manipulation in both electronic and optoelectronic devices. In
particular, spin splitting in heterostructures caused by bulk
inversion asymmetry (BIA) \cite{Dress} and structural inversion
asymmetry (SIA), often called Rashba splitting \cite{R1,R2} has
attracted much attention and understanding various properties and
interactions in these heterostructures is important. Several recent
transport measurements have demonstrated mesoscopic spin-dependent
ballistic transport in InSb-based heterostructures
\cite{Mike1,Mike2}. In addition, the integration of InSb quantum
well (QW) transistors onto silicon substrates has been investigated
recently \cite {Ashley1}. The performance of field effect
transistors (FETs) suitable for digital logic circuits was
demonstrated on material with a buffer just 1.8 ${\mu}$m thick which
is an initial step towards integrating InSb FETs with Si CMOS for
high-speed, energy-efficient logic applications \cite {Ashley2}.


In bulk n-type semiconductors, two spin relaxation mechanisms, the
Dyakonov-Perel (DP) \cite{DP} and Elliot-Yaffet (EY) \cite{EY} are
known to be the dominant relaxation processes. The EY relaxation
process originates from strong mixing of the valence band states and
conduction bands in NGS resulting in non-zero transition rates even
for spin-conserving scattering process. For QWs, the spin relaxation
rate, $\tau{_s}$, in the frame of EY process is modeled as
following: \cite{Tackeuchi3}
\begin {equation}
 \frac{1}{{\tau}{_s}^{EY}}=C_{EY}{\eta}^{2}\biggl(1-\frac{m_e}{m_0}\biggr)^2 \frac{E_{1e}}{E^2_g}k_B T \frac{1}{\tau_p},
\end {equation}\label{Eq.3}
where ${\eta}=\frac{{\Delta}}{E{_g}+{\Delta}}$, ${\Delta}$ is the
spin-orbit splitting of the valence band, $m_e/m_0$ is the electron
effective mass ratio, $E_{1e}$ is the confinement energy for the
lowest electron subband, $E{_g}$ is the band gap, $\tau_p$ is the
momentum relaxation time and $C_{EY}$ is a dimensionless constant
predicted to be nearly unity~\cite{Tackeuchi3}.

In DP mechanism, the lack of inversion symmetry in the presence of a
k-dependent spin-orbit interaction can lift the spin degeneracy even
in the absence of an external magnetic field. In the DP picture, the
spin relaxation rate for QWs is given by~\cite{Tackeuchi1,
Tackeuchi3},

\begin {equation}
 \frac{1}{{\tau}{_s}^{DP}}=C_{DP}\frac{\alpha^2}{2}\frac{E^2_{1e}}{\hbar^2Eg}k_B T \tau_p,
 \end {equation}\label{Eq.4}
where $C_{DP}$ is a dimensionless constant predicted to be 16~\cite{Tackeuchi1, Tackeuchi3}
and $\alpha$ is a parameter characterizing the $k^{3}$ term for conduction band
electrons given by:
\begin {equation}
{\alpha}\approx \frac{4{\eta}}{(3-{\eta})^{1/2}}\frac{m_e}{m_0}
\end {equation}

In the case of the DP process, it would be possible to alter the
spin lifetime with an applied electric field which can modulate the
strength of the spin-orbit coupling via the Rashba effect. In
addition, changing the momentum relaxation time can modulate the
spin relaxation time in both the EY and the DP mechanisms.

In a recent report, the temperature and mobility dependences of the
spin relaxation time in Te-doped InSb/Al$_{0.15}$In$_{0.85}$Sb QWs
have been probed suggesting a fast spin relaxation time of 0.5 ps
\cite{litvinenko}. The anti-localization measurements of the
InSb-based QWs studied in this work have suggested ${\tau}{_s}\sim$
12 ps at temperatures below 10 K \cite{Jean}.
Here we report the dynamics of photo-excited spins in several InSb
based QWs using magneto-optical Kerr effect (MOKE) spectroscopy.
\section{Samples}
We probed relaxation of photo-excited spins in the
Al$_{x}$In$_{1-x}$Sb/InSb QW structures in several optical
excitation regimes. In the first regime the pump/probe pulses were
from a single NIR tunable laser with a maximum fluence of 50
${\mu}$J cm$^{-2}$ on the samples. In the second regime, the pump
excitation was tuned in the mid infrared (MIR) region with a maximum
fluence of 10 mJ cm$^{-2}$ with the probe fixed at 800 nm. We
observed that the photo-excited spin dynamics are strongly
influenced by excitation wavelength, laser fluence, and samples'
growth profiles. Our results are important to develop concepts
toward development of devices employing InSb based heterostructures
and to understand the effect of spin-orbit coupling in the
relaxation dynamics in NGS.
Our InSb square single quantum wells (QWs) were grown on GaAs (001)
substrates by MBE at the University of Oklahoma. The
Al$_{x}$In$_{1-x}$Sb barrier layers are ${\delta}$-doped with Si.
The layers are located either on one side of the QW (asymmetric
sample) or equidistant on both sides of the QW (symmetric sample).
The ${\delta}$ doped layers within the barrier layers are typically
located 70 nm from the well center. The shape and symmetry of the
wells is expected to be determined by whether one or both barriers
are doped. We studied an undoped and five remotely $\delta$- doped
InSb QWs with the electron concentrations in the wells ranging from
${\sim}1-4.4{\times}10^{11}$cm$^{-2}$, where only the ground-state
subband is occupied and the mobility is in the range
${\sim}70,000-100,000$ cm$^{2}$/Vs at 4.2 K. Detailed growth
conditions were described previously \cite{samples1,Giti}. The
characteristics of the samples are summarized in Table I, where
samples S1, S2, S3, A1 and A2 are single modulation doped QWs and M1
is an undoped multi-QW (MQW) structure with 24 wells separated by 50
nm Al$_{x}$In$_{1-x}$Sb barriers.

\begin{table} [htb]
\caption{\it Characteristics of the samples studied in this work.
The densities and mobilities are from the measurements at 4.2 K. In
the doped samples, only the first subband is occupied and the Fermi
levels, $E_{F}$, are with respect to the bottom of conduction band.}
\begin{center}
 \begin{tabular}{|c|c|c|c|c|c|c|}

    \hline
    Sample & Density & Mobility& QW Width & {\it x}& CB1&$E_{F}$ \\
           &  cm$^{-2}$& cm$^2$/Vs &  nm  & {\%} & meV& meV  \\
    \hline
    S1(S769)& 2.0${\times}$10$^{11}$& 100,000 & 30 & 9&14.4&33 \\
    S2(S499) & 1.8${\times}$10$^{11}$ & 135,000 & 30 & 9&14.4&29 \\
    S3(S939) & 4.4${\times}$10$^{11}$ & 96,000 & 11.5 & 15&53&72 \\
    A1(S360)& 2.2${\times}$10$^{11}$ & 73,000 & 30 & 9&14.4&36 \\
    A2(S206)& 1.0${\times}$10$^{11}$ & 70,000& 30 & 7&13.6&16 \\
    M1(S591)& Undoped &  & 30 & 9 & &\\
    \hline

  \end{tabular}
\end{center}
\end{table}

The band offsets in this system has been determined earlier
\cite{offset}. To calculate the interband transition energies we
have used a four-band model described by Bastard \cite {Bastard}
with band-edge masses of 0.0139m$_{0}$, 0.015m$_{0}$, and
0.25m$_{0}$ for the electrons, light holes, and heavy holes,
respectively, where m$_{0}$ is the free electron mass. The results
of the calculations for the interband transitions in our samples at
4.2 K are summarized in Table II. Band-edge effective mass values in
the alloy barrier are considered to change with the band gap
$E_{g}^{x}$ according to the Kane \cite {Kane} model. The energy gap
(in eV) of the alloy at 4.2 K can be calculated from the known
variation of the alloy gap with concentration {\it x}:
$E_g^{x}=E_g^{0}+2.06x$ \cite{Dia98}. In addition, the effect of
strain has been included in determination, $E_{g}^{0}$, the band gap
of the InSb QWs \cite{Dia98-2}. The variation in the bandgap of both
InSb and Al$_{x}$In$_{1-x}$Sb below 77 K is about 3{\%} \cite{Dia98}
resulting in no significant variation in the 2D confinement
potentials and the interband transition energies.
\begin{table} [htb]
\caption{\it Calculated possible interband transition energies for
the samples with different alloy concentrations and well widths at
4.2 K.}
\begin{center}
 \begin{tabular}{|c|c|c|c|c|c|}

    \hline
    Alloy & Well &CB1-HH1&CB2-HH2&CB1-LH1&CB2-LH2 \\
    {\%}&Width &       &       &       &        \\
            & nm & meV($\mu$m) & meV($\mu$m) & meV($\mu$m) &   meV($\mu$m)  \\
    \hline
    7& 30 &261(4.7) &300(4.1)& 291(4.26)& 343(3.6) \\
    9 & 30 & 265(4.68) & 330(3.75)& 302(4.1) & 360(3.45) \\
    15 &11.5& 318(3.9) & 449 (2.76)& 386(3.2) & NA \\
    \hline

  \end{tabular}
\end{center}
\end{table}
\section{Experimental technique}
We probed photo-excited spins in the Al$_{x}$In$_{1-x}$Sb/InSb QW
structures at two optical excitation regimes. The experimental
details of these two regimes are described in this section. Both
degenerate and two-color pump-probe techniques were employed to
study the spin relaxations. The degenerate pump-probe experiments
were performed using a mode-locked Ti-Sapphire laser which produces
tunable radiation from 750 to 850 nm with a repetition rate of 80
MHz, with a maximum fluence of 50 ${\mu}$J cm$^{-2}$ on the samples.
A small portion ($\approx$ 10\%) of the laser beam was split off to
be used as the probe. The pump beam was modulated at frequency of 1
KHz with a mechanical chopper. In this configuration, the excitation
laser can not avoid exciting carriers in the Al$_{x}$In$_{1-x}$Sb
barrier layer. Our earlier time resolved cyclotron resonance
measurements on an undoped InSb MQW demonstrated that exciting the
sample with 800 nm pulses can result in high density of
photo-excited carriers in the wells \cite{Giti-TRCR}.

Our two-color pump-probe measurements were performed using an
optical parametric amplifier (OPA) excited by a Ti-Sapphire chirped
pulse amplifier (CPA) with a repetition rate of 1 KHz. The OPA beam
was used as the MIR pump, with a maximum fluence of 10 mJ cm$^{-2}$
on the samples and a small portion of the CPA (10$^{-5}$) was used
as the probe. In these measurements the pump excitation created
carriers in the barrier layer except when the sample S3 was pumped
at a wavelength of 2.6 ${\mu}$m.

As a result of selection rules for interband transitions,
spin-polarized carriers can be created using circularly polarized
pump beams. The MOKE signal arises from the difference between the
optical coefficients of a material for left and right circularly
polarized light which is proportional to the magnetization M
\cite{Zve}.
\begin {equation}
\eta_{k}+i\theta_{k}=-(\kappa^{+}-\kappa^{-})/(2n(n^{2}-1))
\end {equation}

where $\theta_{k}$ is the Kerr rotation, $\eta_{k}$ is the Kerr
ellipticity, n is the index of refraction, and $\kappa^{+}$ and
$\kappa^{-}$ are the optical susceptibilities of the material for
right (${\sigma}^+$) and left (${\sigma}^-$) circularly polarized
light, respectively. Since $(\kappa^{+}-\kappa^{-})\propto M$, the
MOKE effect can be induced by an external magnetic field, or an
optically or a spontaneous induced magnetization \cite{Zve}. Using a
Wollaston prism, the reflected NIR signal was separated into s- and
p- components which are orthogonal and have equal intensity in the
equilibrium spin density state. In the presence of non-equilibrium
spin polarized carriers, the MOKE signal reflects as an intensity
difference between the s- and p- components of the reflected probe
pulses. The signals were monitored using a Si balanced detector and
were fed into a lock-in amplifier.

\section{Results and Discussion}
Here we report the results of MOKE induced by optical magnetization
to measure spin relaxations. Temporal traces of time resolved MOKE
for the sample A1 at 77 K for ${\sigma}^+$, ${\sigma}^-$, and
linearly polarized lights are shown in Fig. 1. The measurements were
taken under the excitation by NIR radiation at 775 nm with average
power $\sim$ 400 mW (fluence of $\sim$ 50 $\mu$J-cm$^{-2}$)
resulting in a photo-induced carrier density of
$\sim$5$\times$10$^{17}$ cm$^{-3}$. As shown for ${\sigma}^+$, at
timing zero, sample A1 demonstrates a sharp increase in the MOKE
signal followed by a rapid recovery of the signal. The same
measurement for ${\sigma}^-$ shows a similar pattern which is not a
mirror image of the trace shown for ${\sigma}^+$. No residual MOKE
signal was observed for the linear polarization of the pump. The
inset shows the MOKE signal at 4.2 K for ${\sigma}^+$ which exhibits
the same oscillatory pattern. As shown in Fig. 2a for sample A2,
under similar experimental conditions, we observed oscillations at
77 K. The oscillations appear at time delays greater than 10 ps and
do not appear at negative time delays. The oscillatory pattern can
be attributed to precession of {\it z} components of electron spins
at the Fermi level with possible contributions from both BIA and
SIA. A similar damped oscillatory behavior has been observed in a
high-mobility n-doped GaAs/AlGaAs at temperatures below 5 K
\cite{Brand, Stich}.  This effect has been attributed to the
breakdown of collision-dominated regime of spin relaxation. In their
case, the SIA considered to be 10 times less important than the BIA.

\begin{figure} [tbp]
\begin{center}
\includegraphics [scale=0.8]{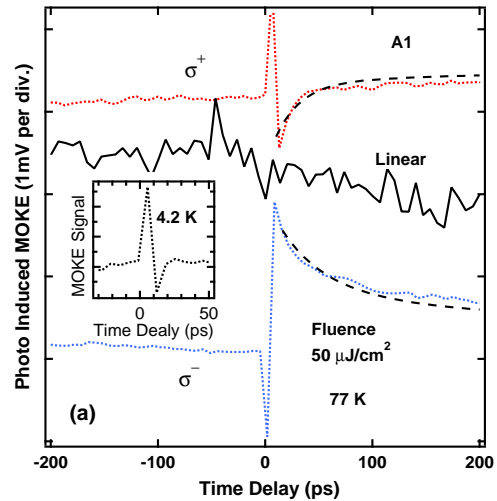}
\caption{ Photo-induced MOKE of sample A1 at 77 K versus time delay
under pumping with circularly and linearly polarized radiations at
775 nm and probing with the same wavelength. The upper trace
represents the photo-induced carrier density. The laser fluence is
estimated to be in the order of 50 ${\mu}$J-cm$^{-2}$. The inset
shows the MOKE measurement for one pump polarization at 4.2 K. The
dashed lines represent exponential fits to the data and for clarity
are shifted slightly.}
\end{center}
\end{figure}

In our case, the calculated potential profile of the conduction band
in a 30 nm wide asymmetric InSb/Al$_{0.09}$In$_{0.91}$Sb QW with
doping density of $\sim$ 2.0${\times}$10$^{11}$ cm$^{-2}$ suggests
an in built electric field of 3.3${\times}$10$^{6}$ V/m
corresponding to an effective magnetic field of $\sim$ 0.3 mT
\cite{Zwad}.  In our samples, the damped oscillations occur at
higher temperatures (77 K or lower) compared to the GaAs/AlGaAs
structures \cite{Brand, Stich}. If our observed oscillations are due
to the spin precession at zero applied magnetic field, then InSb QWs
are potentially useful in spin FETs that operate at higher
temperatures. If we only include the spin-polarized photo-induced
carriers, with a density on the order of $5.0\times10^{9}$
cm$^{-2}$, we obtain oscillation frequency, $\Omega_{(K_{F})}$, at
the Fermi level equal to 0.7 rad/ps corresponding to a period of
$\sim$9 ps. In this scenario, the calculated frequency is the same
order of magnitude as the observed oscillation frequency in our
measurements. The $\Omega_{(K_{F})}$ and the period change to 4
rad/ps and 1.5 ps, respectively, if we include a total density of
$2.0\times10^{11}$ cm$^{-2}$. In these estimations, the Rashba
splitting strength of $\alpha_{R}=1.3\times10^{-9}$ eV-cm was used
\cite{Giti}. Using $\alpha_{R}=1\times10^{-10}$ eV-cm with a total
density of $2.0\times10^{11}$ cm$^{-2}$ results in
$\Omega_{(K_{F})}$ of 0.3 rad/ps, closer to the observed frequency
in our measurements. This smaller order of magnitude for
$\alpha_{R}$ has been calculated for a well width of 30 nm and
electron density of $2.0\times10^{11}$ cm$^{-2}$ in Ref.
\cite{Gilbertson}. Recently the Larmor frequency of bulk InSb has
been measured suggesting a spin precession period of 5-10 ps,
depending on the laser pumping wavelength \cite{InSb-Larmor}.

If the product of $\Omega_{(K_{F})}\tau_{p}\geq1$, spins precess
more than a full cycle before being scattered. On the other hand if
this product is smaller than 1, the individual electron spin can
only precess by some fraction of a full cycle before the momentum
scattering changes the amplitude and direction of the effective
magnetic field \cite{Brand, Stich}. In our case for the measured
$\Omega_{(K_{F})}=0.2$ rad/ps, if we use the value of the momentum
scattering, $\sim$ 0.6 ps, from the measured Hall mobility at 77 K,
a collision-free condition would not be satisfied unless the
momentum scattering time of the photo-induced spin polarized
carriers is significantly larger than the value obtained from the
Hall mobility ($\sim$ 0.6 ps) for the electrons already in the QW.
In the following sections the results of our measurements in
symmetric samples are presented.

Figure 2b demonstrates the typical spin relaxations in the symmetric
samples S1, S2, and the MWQ, M1. For a fluence of about 50
${\mu}$J-cm$^{-2}$ the relaxations in S1, S2 and M1 exhibit some
similar patterns. If we refer to the relaxation time as the time
when the signal at positive time delay reaches approximately to the
same value the negative time delay. The temperature dependence of
relaxations in these structure is rather weak but as shown for S2,
the relaxation time is slightly tunable as a function of the laser
fluence. The lack of oscillations in the MOKE signal of our
symmetric samples could suggest that the BIA is not a dominant
mechanism of the spin splitting in these structures. It has been
suggested and probed by several groups that BIA dominates mainly in
large gap semiconductors \cite{Lommer,Pfeffer1,Pfeffer2,Pfeffer3}.


\begin{figure} [tbp]
\begin{center}
\includegraphics [scale=1.0]{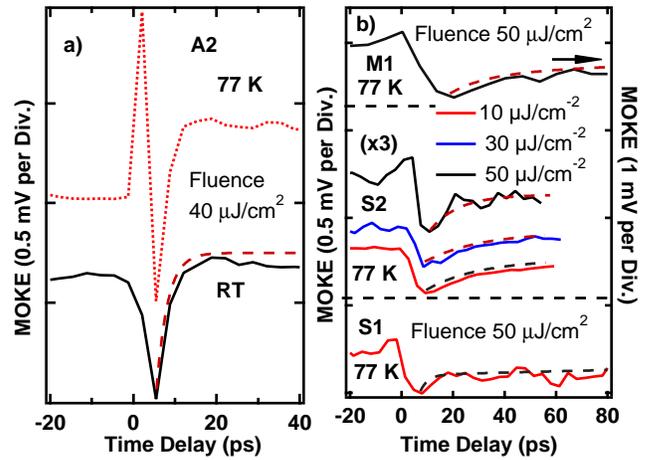}
\caption{a) Photo-induced MOKE of sample A2 at 77 K and room
temperature (RT) versus time delay. Multi-step relaxation disappears
at temperatures above 77 K and the temperature dependence above 77 K
is weak. b) Top trace: MOKE of the undoped sample, M1, at 77 K for
fluence of $\sim$ 50 $\mu$J-cm$^{-2}$. Middle traces: MOKE traces of
the sample S2 at different laser fluences,. Lower trace: MOKE
signals in the sample S1 for a fluence $\sim$ 50 $\mu$J-cm$^{-2}$.
In the samples S1, S2, and M1 the temperature dependence is weak and
data is shown only for the measurements at 77 K. The dashed lines
represent exponential fits to the data and for clarity are shifted
slightly.}
\end{center}
\end{figure}

In the second regime  probed by our measurements, we used a
two-color MOKE scheme, using MIR pump pulses fixed at 2 $\mu$m and
NIR probe pulses fixed at 800 nm. For the samples studied here 2
$\mu m$ (620 meV) pump pulses can not avoid exciting the carriers in
the Al$_{x}$In$_{1-x}$Sb barrier layers, but the significance of
this scheme was that it allowed probing the relaxations at a higher
fluence compared to the degenerate regime. Examples of the
measurements are shown in Fig. 3a and 3b for two different samples
(A1 and S1) at RT. In this regime the photo-induced spin relaxations
show exponential decays faster by an order of magnitude compared to
the lower fluence regimes shown in Fig 2. This can be explained
using the EY mechanism where the spin relaxation time is directly
proportional to momentum scattering time. At higher excitation
fluences the momentum scattering time is expected to be shorter (due
to the existence of high photo-induced carrier density) and the
observation of a faster spin relaxation is expected for the EY
mechanism.

Figure 3c shows the spin relaxation traces for the sample S3 at
several temperatures in the high fluence regime. Compared to the
samples A1 and S1, the sample S3 shows a longer spin relaxation time
in the high fluence regime. In the case of S3, using 2.6 $\mu$m
radiation (477 meV, this photon energy is close to the HH2-CB2
transition in this sample) for the pump, it is possible to excite
carriers only in the InSb well and not in the
Al$_{0.15}$In$_{0.85}$Sb barrier layer. This fact is supported by
earlier measurements that determined the concentration and
temperature dependence of the fundamental energy gap in
Al$_{x}$In$_{1-x}$Sb \cite{Dia98}. In addition, as shown recently
for bulk InSb at RT \cite{MurdinFEL}, a large density of photoholes
can reduce the momentum relaxation time, $(\tau_p)$, of the
electrons and therefore if the DP process is dominant, an increase
in the spin lifetime can be observed. In bulk n-InSb this effect was
found to increase the spin lifetime by a factor of 2 to 3 , changing
the spin relaxation from 14 to 38 ps. As shown in Fig. 3c at 77 K
for several laser fluences, we observe modification of the
relaxation patterns by the density of photo-induced carriers. The
effect of pump fluence and therefore initial spin polarization in a
high mobility GaAs/AlGaAs single QW has been probed, suggesting a
similar increase in spin relaxation time by increasing the initial
spin polarization \cite {Stich}. In our samples, we only see this
effect when we selectively pump the InSb well layer and avoid the
barrier materials.

\begin{figure} [tbp]
\begin{center}
\includegraphics [scale=0.9]{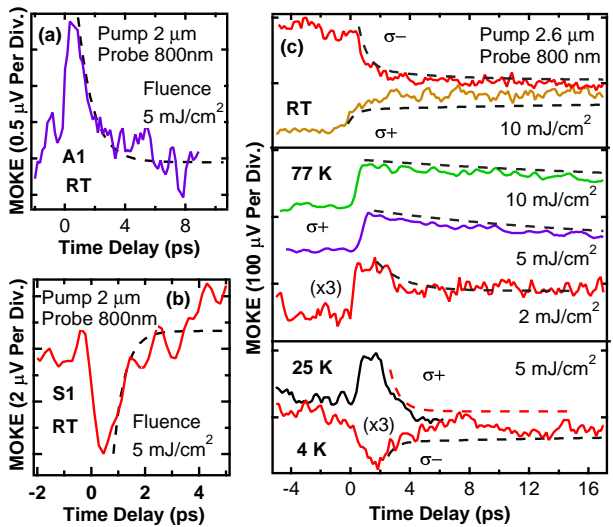}
\caption{a) and b) show examples of the two-color MOKE measurements
in high fluence regime.  c) MOKE traces of sample S3 at different
laser fluences and temperatures.  At RT we were able to observe MOKE
at the highest available fluence (10 mJ-cm$^{-2}$) where the
relaxations show a step pattern rather than an exponential decay.
The dashed lines represent exponential fits to the data and for
clarity are shifted slightly.}
\end{center}
\end{figure}

In general, we believe that some of the observed differences could
be also due to the differences in the growth profiles. For example,
S3 (the 11.5 wide QW) has 13 nm of InSb cap layer and 139.5 nm of
Al$_{x}$In$_{1-x}$Sb layer above the well, whereas the 30 nm QWs,
except S2, has a 10 nm thick InSb cap layer. S2 has a 20 nm thick
InSb cap layer. In addition, samples A1, A2, S1 and S2 have 25 nm,
60 nm, 170 nm and 130 nm of Al$_{x}$In$_{1-x}$Sb above the InSb QW
layers, respectively. Sample M1 which is a MQW, has a 10 nm thick
InSb cap layer with 50 nm thick barriers above and below each well.
The absorption length in these heterostructures (in the wavelength
region used in this study) is about 0.5 $\mu$m ; therefore, the
regions where the photo-generated carriers are created can be
different in each sample. In our measurements, the best case to
compare the experimental data with the DP and EY models, is the case
in Fig. 3c. In this sample (S3) using 2.6 $\mu$m, we have avoided
the barrier material. Using the momentum scattering time from the
measured Hall mobility, the calculated spin relaxation, $\tau_s$,
according to the EY and DP models are summarized in Table III. It
has been theoretically predicted that for III-V bulk semiconductors,
EY mechanism is dominant at very low temperatures (T$<$6 K)
\cite{Song}. However, it has been reported that a cross-over to the
DP mechanism can occur even at RT in low mobility ($\sim$10,000
cm$^2$/Vs) InSb QW samples \cite {Litvinenko_NJP}. In our case the
EY model provides a better fit for the RT observation and the DP for
the measurements at 4.2 K.

\begin{table}[htb]
\caption{\it Calculated spin relaxation times $\tau_s$ with
$C_{DP}=16$ and $C_{EY}=1$ for sample S3 using Eq. 1 and Eq. 2,
compared to the experimental observations at several temperatures
with a significant variation in the relaxation times.
  }
  \centering

  \begin{tabular}{|c|c|c|c|c|}
      \hline
  Sample &T(K)& $\tau_s(DP)$ & $\tau_s(EY)$ & $\tau_s$\\
  &      &  (ps) theory   &    (ps) theory  & (ps) experiment\\
   \hline

 S3&4.2&7&3827&$\sim$8\\

      &25&1.3&535&$\sim$6\\

      & 77&0.6&124&  $>$16\\

      & 300 &0.4&12& $\gg$ 16 \\

     \hline
 \end{tabular}
\end{table}


\section{Conclusions}
We report spin relaxation measurements in a series of InSb based
QWs. In the low fluence regime ($\sim$ 50 $\mu$J-cm$^{-2}$), an
oscillatory pattern was observed in samples with asymmetric doping
profiles at low temperatures ($\leq$ 77K) which can be attributed to
a quasi-collision-free spin relaxation regime. Probing this effect
further by gating the samples or by adjusting the asymmetric doping
level to alter the built-in electric field and therefore tuning the
effective magnetic field, can provide more insight to this
observation. In addition, the actual value of $\alpha_{R}$ (the SIA
factor or the Rashba coefficient) which controls the precession
frequency of the photoexcited carriers is a crucial factor to model
the observations.  In a high fluence regime, using a two-color setup
and not avoiding the barrier materials, we observed faster spin
relaxations expected from the EY mechanism. By selectively pumping
the well of sample S3, we observed spin relaxation times which were
different from non-selective pumping schemes. This fact might
suggest that the spin relaxation observed in the non-selective
pumping can be influenced by the barrier materials. We will be
extending our measurements to probe samples with higher alloy
concentrations when they become available and to probe the spin
relaxations using the differential transmission technique in MIR.

{\bf Acknowledgment:} This work has been supported by
NSF-DMR-0507866, Jeffress-J748, Advance-VT, AFOSR Young Investigator
Program-06NE231, NSF-DMR-0510056 and DMR-0520550.



\newpage

\end{document}